# The effect of duty cycle on electron transmission through a graphene electrostatic barrier

by

R. Biswas[a], S. Mukhopadhyay[b] and C. Sinha[c]

[a]Department of Physics, Rammoham College, Kolkata-700009, W.B.

[b]Department of Physics, Chunaram Gobinda Mem. Govt. College, Susunia, Purulia-723131, W.B.

[c]Indian Association for the Cultivation of Science, Jadavpur, Kolkata-700032, W.B.

**Abstract:**

We investigated theoretically the transmission properties of Dirac Fermions tunneling through a periodically (sinusoidal and rectangular) driven electrostatic barrier in Monolayer graphene. For the time harmonic potential with moderate to high $\alpha \,(= V_0/\hbar\omega)$ the central Floquet band is found to be almost cloaked for the Klein transmitted electron in contrast to electron at higher grazing incidences. As a time periodic drive, we mainly focused on the use of rectangular wave electric signal to modulate the transparency of the barrier. It is noted that the asymmetric Fano resonance, a characteristic feature of photon assisted tunneling, is more likely to occur for rectangular drive in contrast to the harmonic one. The height of the modulating potential is particularly responsible for the dressing effect of the barrier. The position and nature of the FR can be tailored by changing the height and frequency of the rectangular drive. Moreover, the duty cycle of the driving potential turns out to be an important controlling parameter for the transmission process. Thus, the rectangular modulation plays an important role for the occurrence and detection of the Fano resonances which is vital for the use of graphene nanostructure in the field of detectors, sensors, modulators etc. The present findings attempt for the first time, to realize the effect of duty cycle on the quantum interference in semiconductor nanostructures.

1. Introduction

The experimental realization of graphene by Novoselov et. al.[1] in 2004, one of the major breakthrough in science, remains among the most important discoveries in the field of Science and technology. This discovery provided an immense boost in the experimental realization of various 2D materials [2, 3] and created a new dimension to material research and applications. The unique and exotic electronic, optical, mechanical, thermal and chemical properties, along with their successful applications, make graphene a fascinating and exceptional carbon material that has created a lot of interest both theoretically as well as experimentally during the last two decades. Graphene, a dream material, exhibits electron mobility 140 times higher than Si and a strength 200 times that of steel. As regards mechanical properties, very recent applications [4] proposed that graphene, when mixed with concrete, leads to better bonding at the microscopic level, yielding a stronger (strength increases by 30%), more durable and more corrosion resistance material. Lots of theoretical and experimental research works have been carried out [5] in the burning field of nano electronics (e.g., thin film transistors, quantum dot devices, etc.), supercapacitors, fuel cells, batteries, photovoltaics, catalysis, separation and storage, gas absorption, sensors, etc. Further, by virtue of its high surface to volume ratio, the material graphene offers itself as a very promising and efficient candidate to be used in batteries and supercapacitors. Graphene technology could also be utilized for rapid detection of infections and thereby act as the basis for new medical diagnostics [6, 7]. So far as the electronic property is concerned, the observation of the Klein transmission [8, 9] and the anomalous quantum Hall effect [10] arise due to the characteristic energy band structure described by the massless Dirac Weyl fermions distinguishing graphene from conventional 2D materials. Following the seminal work of "Chiral tunneling of Dirac fermions in monolayer graphene" by Katsnelson and Novoselov in 2006 [11], series of works were carried out on the tunneling transport through the single and multiple electrostatic barriers in graphene. The Klein transmission, the Febry Perrot resonances and the transmission angular anisotropy are some of the characteristic features of the graphene based tunneling nanostructures [11,12].

The external control of the electronic transport plays a vital role for the exploitation of the nanostructures towards the digital applications. Greater the number of controlling parameters, greater is the usefulness of the external means. In this regard, applications of static electric and magnetic fields are well known in the literature. The tailoring of the electronic transport has gained much momentum considering the electron photon interaction in low dimensions. In particular, the recent development of microwave technology and laser physics has created much interest in this direction. The application of laser on 2D micro and nanostructures has led to significant modifications in the physical properties of the system [13-17]. It was shown by Calvo et. al. [14] that by the application of a circularly polarized light, the energy dispersion of the Dirac fermions in graphene is modified and a finite band gap is created depending on the frequency and intensity of the laser field. It results to the suppression of the Klein transmission through a graphene electrostatic barrier. On the other hand, an anisotropy factor can be introduced in the energy dispersion of graphene by the application of a linearly polarized off-resonant dressing field leading to an asymmetric off-normal Klein paradox [15] and suppressed Klein tunneling [16, 17] in the transmission spectrum.  Theoretical study of the photon assisted tunneling dates back to 1963 when Tien and Gordon [18] justified the experimental observation of photon exchanged tunneling transport experiment on superconducting film in presence of the microwave field. Following the above mentioned photon assisted tunneling process lots of efforts [19, 20] were carried on to study the effects of oscillating time dependent potential on the transmission of Schrödinger electron through a potential barrier/well structures. Tunneling of Dirac fermions in presence of oscillating potential were first reported by Trauzettel et. al. [21] to probe energy dependent transmission and to the observability of zitterbewegung in graphene nanostructure. Recently the effect of harmonic potential on the Klein transmission [22] and Febry Perot resonances [23] are studied theoretically using the non-perturbative theory of Floquet formalism. It has been proposed that in presence of time harmonic potential discrete asymmetric Fano resonances (FR) may appear due to the effect of quantum interference between the discrete and continuum states [24-28].

Uptil now, the application of the time dependent potential to the tunneling nanostructures is limited to sinusoidal nature only [18-36]. However, in view of the advancement in the electronic technology it has now become feasible to construct non-sinusoidal generators that are capable to generate electric pulses other than the sinusoidal nature. Different types of the non-sinusoidal time dependent electrical signals are namely square wave, triangular wave, sawtooth wave pulses etc. In particular, applications of square wave (SW) pulses have created a revolution in the nano-scale industry, specifically in the production of the digital signal that lies at the heart of the memory devices [37]. The SW signal consists of only two levels, making it easy to generate and analyze. The high and low states of the square wave pulse are termed as the '1' and '0' states respectively that are the building blocks of the signal in the digital world [38]. The SW signal finds extensive use in digital systems, telecommunications and switching power supplies etc. It is advantageous in digital systems for quick switching between high and low states. They could be used as clock signals, because their fast transitions are suitable for triggering synchronous logic circuits at precisely determined intervals [38]. In spite of the wide range of applications of the square and rectangular wave pulses in the digital industry, the application of the non-sinusoidal pulses to the electronic transport in nano scale tunneling devices is not yet reported in the literature.

Motivated by the aforesaid range of applications of the non-sinusoidal wave pulses, the present work addresses theoretically the effect of time varying rectangular wave potentials (RWP) on the tunneling transport of the Dirac fermions in graphene nanostructures. The application of the RWP virtually modifies the time dependence of the external control in such a way, that it becomes equivalent to a sinusoidal potential at a given frequency along with its higher harmonics. As a result, it is expected that the presence of harmonics will modify the transmission characteristics in a different way as compared to the application of a monochromatic sinusoidal signal. As is known, the square wave pulse is a special type of rectangular wave pulse with the duty cycle 50% and it is possible to vary the duty cycle by changing the electronic components of the generator. The average output increases or decreases depending on the increase or decrease of the duty cycle. So far as the application is concerned, the effect of duty cycle has been studied to explain the morphology, composition and chemical properties of coating deposited by pulsed arc ion plating [39]. Thus, the use of rectangular time varying potential provides us a new controlling parameter D, the duty cycle, other than the amplitude and frequency, thereby making it superior [40, 41] over the sinusoidal potential to manipulate the transmission in nano-structure based *n-p-n* systems.

## 2. Models and method

The tunneling potential profile considered in the present report is shown in Fig. 1(a), where along the x-direction the graphene sheet is divided into three regions forming a 2D model of a n-p-n transistor. The system is uniform along the y-direction. Regions I and III act as the source and drain terminals respectively, whereas the external time dependent potential is applied in region II (Fig.1(b)) by the application of a top gate, leading to the formation of an oscillating barrier region in the model.

In carbon monolayer the charge carrier obeys the well-known linear dispersion given by the Dirac like Hamiltonian ($H_0$) and in presence of the external potential U(x,t) the full Hamiltonian reads as

$$H = H_0 + U(x,t) \quad \ldots\ldots\ldots\ldots\ldots\ldots\ldots\ldots\ldots..(1)$$

where $H_0 = \hbar v_F(\sigma_x k_x + \sigma_y k_y)$ ; $\quad\ldots\ldots\ldots\ldots\quad$ (2)

$v_F$ being the Fermi velocity, $\sigma_x$ and $\sigma_y$ are the component of Pauli matrices, $k_x$ and $k_y$ are the components of wave vector along the x and the y directions respectively. The time varying potential profile is spatially non-uniform along the x-direction and is given by

$$U(x,t) = V_z + V(t) \text{ for } 0 < x < L \quad (3)$$
$$= 0 \quad \text{elsewhere.}$$

'L' corresponds to the width of the barrier and $V_z$ being the static barrier height. The schematic diagram for the proposed structure is shown in Fig. 1(c). The constant barrier height $V_z$ can be tuned by the back gate voltage from a DC source ($V_g$, as shown in Fig.1(c)) whereas the time dependent part V(t) can be controlled by applying a top gate connected to a multivibrator, an AC source ($V_{ac}$) of rectangular wave generator. Here the time dependent part $V(t)$ is even periodic function in character and is given by the rectangular wave nature such that the potential has a constant value $V_0$ for the time interval -$T_p$/2 and $T_p$/2 and zero for the rest of the cycle -$T$/2 to $T$/2 and so on (shown in Fig.1(b)), i.e.,

$$V(t) = V_0 \text{ for } -T_p/2 < t < T_p/2 \quad (4)$$
$$= 0 \quad \text{for rest of the cycle.}$$

$V_0$ and $T$ being the height and the period of the rectangular wave potential respectively, whereas $T_p$ being the width of the pulse. The ratio $T_p/T$ is known as the duty cycle (D) of the periodic pulse. For D = 50%, i.e., for $T_p=T/2$, the signal is known as a square wave. It is well-known that a periodic function $X(t)$, having period $T$, can be expanded by the Fourier series as $X(t) = a_0 + \sum_{n=1}^{\infty}[a_n Cos(n\omega t) + b_n Sin(n\omega t)]$ where angular frequency $\omega = 2\pi/T$, $a_0 = \frac{1}{T}\int_{-T/2}^{T/2} X(t)dt$, $a_n = \frac{2}{T}\int_{-T/2}^{T/2} X(t)Cos(n\omega t)dt$ and $b_n = \frac{2}{T}\int_{-T/2}^{T/2} X(t)Sin(n\omega t)dt$. Thus for the square wave signal given by eqn.(4) (under the condition $T_p = T/2$) one can write

$$V(t) = \frac{V_0}{2} + \frac{2V_0}{\pi} Cos(\omega t) - \frac{2V_0}{3\pi} Cos(3\omega t) + \frac{2V_0}{5\pi} Cos(5\omega t) - \ldots \ldots \ldots \quad (5)$$

The above expression shows that the square wave of peak height $V_0$ and time period $T$ can be decomposed into a static part of magnitude $V_0/2$, a sinusoidal term of amplitude $\frac{2V_0}{\pi}$ and a series of odd harmonics with gradually decreasing amplitude corresponding to the factor 1/3, 1/5, 1/7 etc…..of the fundamental one.

Now to solve the problem of transmission we apply the Floquet formalism where one have to find solutions of the time dependent Wyel equation of the form $[H_0 + U(x,t)]\Psi(x,y,t) = i\hbar \frac{\partial}{\partial t}\Psi(x,y,t)$ with $U(x,t)$ given by eqns.(2)-(4) according to the regions I, II and III shown in Figs.1(a) and 1(b). The Floquet problem has been solved following similar procedure as mentioned in previous papers [14, 17, 20, 26, 42,43].

For an electron of wave vector $k_1^0$ incident on the barrier with energy E (corresponding to Floquet central band energy) may be reflected (in region-I) and transmitted (in region-III) through different Floquet side bands having energies $E \pm n\hbar\omega$ due to exchange of photons with the oscillating potential inside the barrier. The wave functions in the regions I and III can be written as

$$\begin{pmatrix} \varphi_a^I(x,y,t) \\ \varphi_b^I(x,y,t) \end{pmatrix} = A_0 \begin{pmatrix} 1 \\ \frac{k_1^0+ik_y}{E} \end{pmatrix} e^{ik_1^0 x + ik_y y} e^{-iEt} + \sum_m B_m \begin{pmatrix} 1 \\ \frac{-k_x^m+ik_y}{E+m\omega} \end{pmatrix} e^{-ik_x^m x + ik_y y} e^{-i(E+m\omega)t}$$

and $\begin{pmatrix} \varphi_a^{III}(x,y,t) \\ \varphi_b^{III}(x,y,t) \end{pmatrix} = \sum_m F_m \begin{pmatrix} 1 \\ \frac{k_x^m+ik_y}{E+m\omega} \end{pmatrix} e^{ik_x^m x + ik_y y} e^{-i(E+m\omega)t}$  (6)

where $(k_x^m)^2 = (E + m\omega)^2 - k_y^2$. $A_0$, $B_m$ and $F_m$ are the amplitudes of the incident wave, reflected wave of m-th Floquet side band and transmitted wave of m-th side band respectively. In the region II, the solution for the case of sinusoidal oscillation is given by

$$\begin{pmatrix} \varphi_a^{II}(x,y,t) \\ \varphi_b^{II}(x,y,t) \end{pmatrix} = \sum_{m,n} C_m \begin{pmatrix} 1 \\ \frac{q_x^m + ik_y}{E - V_z + m\omega} \end{pmatrix} e^{iq_x^m x + ik_y y} e^{-i(E+n\omega)t} J_{n-m}(\alpha) +$$

$$\sum_{n,m} D_m \begin{pmatrix} 1 \\ \frac{-q_x^m + ik_y}{E - V_z + m\omega} \end{pmatrix} e^{-iq_x^m x + ik_y y} e^{-i(E+n\omega)t} J_{n-m}(\alpha) \qquad (7)$$

where $C_m$ and $D_m$ are constant coefficients and $J_{n-m}(\alpha = V_0/\omega)$ is the Bessel function that will be modified in case of the rectangular waveform. Finally, matching the wavefunction at the interfaces $x = 0$ and $x = L$ one can find out the transmission coefficient in the $m^{\text{th}}$ Floquet side band state, given by $T_m = \frac{\cos\theta_m}{\cos\theta_0} \left|\frac{F_m}{A_0}\right|^2$, where $\theta_0$ and $\theta_m$ are the angles corresponding to the incident state and $m^{\text{th}}$ transmitted/reflected side band respectively.

Now utilising the above results for the transmission coefficients one can calculate the zero temperature conductance by applying the Landauer-Buttiker formalism [44, 45] using the relation

$$\sigma = \sigma_0 \int_{-\frac{\pi}{2}}^{\frac{\pi}{2}} T(\theta, E, V_0) \cos\theta \, d\theta \qquad (8)$$

where $T(\theta, E, V_0)$, the total transmission summed over all the Floquet sidebands, is given by $T(\theta, E, V_0) = \sum_m T_m$ and $\sigma_0 = \frac{gEL_y}{\pi}$ (in units of $e^2/h$), $g$ is the valley degeneracy, $L_y$ is the length of the structure along the y-direction and $E$ is the Fermi energy.

## 3. Results and Discussion

Let us first start our discussions about the numerical results (Figs.2(a)-2(d)) for the case of a time dependent harmonic potential (TDHP) applied on a graphene based static electrostatic barrier [22]. Here, the length, energy and frequency are expressed in units of 81.13nm, 8.113mV and 12.325x10$^{12}$Hz respectively, unless otherwise specified. Fig.2(a) represents the variation of transmission coefficients ($T_m$ for the m$^{th}$ Floquet sideband and T for summed over all sidebands) with the strength of the oscillating potential $\alpha$ $\left(=\frac{V_0}{\omega}, V_0 \text{ and } \omega \text{ being the amplitude and frequency of sinusoidal potential}\right)$ for the case of incident energy E=10.107, incident angle $\theta=30^0$, barrier height $V_z$= 24.65, frequency $\omega$=0.406 and barrier width L=1.233 [22]. The figure reveals that the total transmission under sinusoidal drive is oscillatory in nature (for the present parameter, the magnitude always higher than the static barrier transmission, T=0.51555). This implies that the barrier transparency may increase with the application of a sinusoidal potential. Further, the characteristic oscillatory behavior of T with respect to $\alpha$ is a manifestation of the Bessel function appeared in the virtual photon exchanged electronic states inside the barrier. So far as the side band transmission is concerned, it may be pointed out that for the lower value of $\alpha$ (i.e., lower amplitude or higher frequency of the sinusoidal modulation) the photon exchanged transmission probabilities are negligible (e.g., less than 10% for $\alpha$<2). However, with the increase in $\alpha$, the photon assisted channels become more and more accessible thereby making the total transmission higher than the static barrier limit. On the contrary, the characteristic Klein transmission (with T = 1) is preserved for normal incidence (not shown) even in the presence of the oscillating potential [22]. Therefore, for glancing incidence, the oscillating potential not only redistributes the transmitted electrons among different Floquet sidebands but may increase the total probability of barrier penetration (unlike the normal incidence, for which the TDHP only redistributes the electrons with the total transmission always summed up to the static value of unity).

In order to study the angular dependence of the modulated transmission, we have plotted in Fig.2(b) the total transmission (summed over all Floquet channels) for four different values of α (e.g., 0, 0.5, 2 and 5 corresponding to $V_0$ =0, 0.203, 0.812 and 2.03 respectively). The angular transmission profile of the static barrier is an even function of the angle of incidence and exhibits some sharp resonant peaks known as Febry Perot resonances (FPR) [11,12]. These large angle peaks are obtained when the x-component of wave vector inside the barrier satisfies the condition $k_2 = \frac{p\pi}{L}$, where p being an integer and L being the width of the barrier [11]. Under the application of the TDHP, the directional symmetry of the angular transmission profile ($T_\varphi = T_{-\varphi}$) in static barrier remains protected. For lower values of α, the transmission signature maintains the static barrier character except for a significant modulation of the height of the resonant peaks, not protected by the chirality (FPR). The suppression is more prominent for higher glancing incidence. On the other hand, for higher values of α, extra resonance peak arises in the transmission profile. This may be attributed to the increase in oscillatory nature of the Bessel function with the increase in the order m (the Floquet side band index). In fact, the time average behavior is introduced through the appearance of the Bessel function in this formalism. In effect, the charge carriers are only redistributed among various Floquet transmission and reflection side bands where the probability of distribution depends on the amplitude and frequency of the oscillating potential, not on time. Thus, the oscillatory potential modifies the transmission profile of the Dirac fermions unless they are protected by chirality. The Klein transmission under oscillating potential bears some characteristic features as follows: Firstly, the property of Klein transmission is preserved even in presence of the oscillating potential. Secondly, the sideband transmission probability depends significantly on the number of photon absorption or emission inside the barrier, although for a given photon number the absorption and emission channels have the same probability ($T_m = T_{-m},$), a characteristic feature of the dynamic KT (the photon assisted

KT). Thirdly, for higher α, although the central band remains the most probable transmission channel at higher grazing incidence, it becomes the lowest probable one under the Klein transmission. Now due to the conservation of the transverse momentum $(k_y = E_m Sin\theta_m)$ all the Klien electrons $(k_y = 0)$ in different Floquet channel ($E_m$) will transmit normal to the interface, in contrast to the glancing incidence where no electron will transmit normal to the interface. Under oscillating potential, the process of electron redistribution is highly directional in nature. For higher amplitude, the transmission through the central band is almost cloaked for normal incidence, as can be seen from Fig. 2(c). The situation can be coined as Cloaked Klein Floquet tunneling. But the situation is just reverse for higher glancing incidence. T$_0$ remains the highest probable transmission for the FPR peaks (the non-Klein resonances) and photon absorbed transmissions are more dominant than that for the photon emitted transmission channels (vide Figs. 2(c) and 2(d)). For glancing incidence, lower energy Floquet electron will scatter more than its higher energy counterpart. Further, the transmitted electron undergoing zero photon process moves parallel to the incident electron but those for photon transfer process, propagates along different directions depending on their energies (or in other words depending on the number of photon transfer). Thus, the oscillating barrier behaves as a system with multiple refractive indexes, so far as the electron optics is concerned [46] and the system can act as a diverging lens for electrons at glancing incidences. It may be noted from Fig.2(d) that the Non-Klein transmission resonances are mostly destroyed for the photon emitted transmission processes. Thus, we can conclude that, after crossing the time modulated electrostatic barrier, the Klein electrons are grouped along the energy scale whereas the non-Klein electrons are grouped energetically as well as directionally due to the effect of the time harmonic potential.

After discussing some important features of electron transmission through a time dependent harmonic potential (TDHP) let us now focus on the results of the actual problem of time dependent square wave potential (TDSP) applied on a graphene electrostatic barrier. We have applied the previously adopted transfer matrix method [42, 43], widely used for tunneling problem, to calculate the transmission coefficients for the central and different Floquet sidebands. For direct comparison between the results from harmonic and square wave potentials, the functional form of the sinusoidal potential has been changed accordingly. From Fig.3(a) it is clear that for square wave modulation also, the angular symmetry ($T_\varphi = T_{-\varphi}$) of the transmission spectra is maintained as in the previous cases of TDHP [22]. Not only that, the Klein transmission for the field free electrostatic barrier remains almost unaffected. However, other non-Klein resonances (i.e., FPR with T=1) are found to be suppressed significantly in presence of the TDSP modulation. In comparison to the static barrier, the transmission is more oscillatory in nature, even more than the sinusoidal modulation. The main feature of the square wave modulation is the appearance of an asymmetric Fano line shape (around θ~71.5$^0$) instead of a sharp resonance (FPR) with perfect transmission at the large angle. Although the Fano resonance (the quantum interference between discrete and continuum pathways) is the characteristic feature of photon assisted tunneling, it was not observed for the case of a sinusoidal time modulation (TDHP). Strictly speaking, this conclusion might not be general. In fact, the appearance of the FR depends on a number of factors, e.g., the position of quasi-bound state of the static barrier, number of photon exchange required to satisfy the energy conservation relation etc. The absence of FR for the TDHP could be due to the fact that in this case, in order to obey the energy conservation, we need to exchange at least three photons (which is less probable) in contrast to the single photon process required for the TDSP.

In order to study the effect of amplitude of the TDSP, we have plotted in Fig.3(b), the results for three different heights ($V_0$) of the TDSP (e.g., 0.1, 0.5 and 0.8). For lower amplitude ($V_0$ ~ 0.1) of the TDSP, lower angle transmissions and FPR remain almost unaffected in magnitude (only a shift in angular position). While for higher angle, the resonance of the static barrier disappears and a narrow FR is noted in the transmission spectra. With the increase in height of TDSP, the position of the FR gradually shifted to the higher incident angle and the width of the FR increases so that the asymmetry is clearly visible (Fig.3b). Due to the presence of the TDSP the static barrier is dressed such that the position of the field free quasi-bound state is now shifted with respect to their field free position and the amount of shift is dependent on the magnitude of dressing, i.e. on the height of the TDSP. Thus, the shifting of FR with respect to the change in the height of the TDSP may be due to the dressing effect of the barrier.

To study the effect of frequency of the TDSP on the transmission we plotted in Fig.3(c) the total transmission for three different frequencies, e.g., 0.2, 0.4 and 1.0. It may be noted that the Klein as well as low angle transmissions remain unaltered for all frequencies, although remarkable changes are noted in the FPRs (for higher glancing incidences). The effect of TDSP modulation on the non-Klein resonances are much more pronounced at lower frequencies (e.g., $\omega = 0.2$) as mentioned earlier for TDHP. The extreme high angle resonant peak almost disappears, instead a sharp asymmetric FR is noted within the angular range $60^0$ to $90^0$. The position of the FR moves towards lower angle and the Fano minimum becomes more prominent with increasing frequency (where for low frequency the FR is line type in nature) of the TDSP. As was mentioned earlier the existence of the FR is related to the presence of quasi-bound states inside the nanostructure. In graphene nanostructure for normal incidence the solution is oscillatory for all the three regions. Whereas for glancing incidence, the y-component of momentum acts as barrier in regions I and III, so that quasi-bound states may exist in the region II for incident energy in the range between zero and $|k_y|$. As $|k_y|$ increases with the increase in the magnitude of the incident angle (keeping E fixed), the possibility for the presence of quasi-bound state also increases for higher glancing incidence. It is well known that the appearance of the FR at energy '$E$' is determined by the energy condition $E_b = E \pm n\omega$, $E_b$ being the quasi-bound state energy. When '$E$' is far away from $E_b$, then for a particular '$\omega$', $n$ should be large (and hence less probable) for satisfying the above relation in case of TDHP compared to TDSP. Thus, due to the presence of higher order harmonic terms in the TDSP, the probability of satisfying the above energy condition is greater for TDSP than the TDHP. Therefore, the appearance of the FR is more probable for TDSP than the TDHP modulation. Hence, the Fano dynamics may be easily expected in the case of the square wave modulation.

As we have already mentioned that the use of the rectangular potential modulation in the time domain possesses an extra parameter '$D$', the duty cycle (DC), for tailoring the electron transmission in the nanostructures. As already mentioned, the duty cycle defines the fraction of time over which the time dependent potential is nonvanishing over a complete cycle. As for example, D=25% implies that during one complete period T the potential is nonvanishing for the time T/4 and zero over the rest 3T/4.     Thus, to study the effect of DC on the transmission we have plotted in Figs. 4 (a) and 4(b) the total transmission for four different values of the DC, e.g., 10%, 25%, 50% and 75%. It is customary to mention here that with the decrease in DC the effective time of modulation over a period of the time varying potential on the transmission decreases. It may be noted from Fig.4(a) that a very low width of the rectangular pulse has almost no effect on the transmission for the low and moderate values of the incident angle, as expected, since the interaction time is very small. There is a systematic change in the position of the intermediate angle FPR towards higher incident angle with increase in DC. The peak height decreases with the increase in DC, reaches a minimum value at DC equal to 50% after which height increases again. Further, the angular width of the Klein resonance increases with the increase in DC. Interestingly, there is a particular glancing incidence for which the total transmission is independent of the duty cycle of the oscillating potential. The transmission profile is significantly modified for the higher grazing incidence (clearly viewed from Fig.4(b)) where the FPR under static barrier almost disappears with the appearance of discrete line type resonances (extremely sharp FRs). The number, position and the peak to dip ratio of the FRs are highly dependent on the DC. With the increase in DC, the position of Fano resonances (e.g., FR1 in Fig.4(b)) gradually move towards higher angle with the corresponding increase in the P-to-D ratio. It implies that the dressing effect of the barrier (by dressing we mean the change in static barrier height in presence of the oscillating potential) is highly dependent on the duty cycle of the time periodic rectangular potential. Thus, we find that DC plays a vital role for detectable FR in the transmission spectrum.

To compare and verify the aforesaid theoretical transmission properties (the noted FR) with experiment, it is necessary to measure the differential conductance, a technique in differential conductance spectroscopy. Whereas, in the standard experiment, the directly measurable quantity is the conductance, that involves exhaustive numerical computations for theoretical predictions, particularly when the transmission spectra exhibit Fano type extremely sharp resonances. Further, since the calculation involves extra integration, it requires more computation time and becomes increasingly difficult to overcome the convergence problem. In spite of that, to study the effect of duty cycle, a new parameter for controlling the transmission features, we have presented in Fig.5 the result of conductance with respect to the variation of the parameter $V_0$ for three different values of the duty cycle, e.g., 25%, 50% and 75%. For $V_0 = 0$ the current is carried by the elastic channel. For extremely low values of $V_0$, i.e., when the inelastic channels start contributing the transmission processes, the conductance initially decreases from the static barrier ($V_0=0$) value. Afterwards, it gradually increases and finally exhibits oscillatory behavior for higher values of $V_0$. The oscillation is remarkable and interesting for lower value of DC, e.g., 25% as noted from Fig.5. $\sigma/\sigma_0$ attains a maximum value when $\alpha \sim 5$. We thus note that the duty cycle plays a vital role for manipulating the angle averaged conductance in a graphene electrostatic barrier.

## 4. Conclusion

The Floquet transmission spectra for a graphene based n-p-n structure are reported theoretically for the first time in presence of a time dependent rectangular wave potential as well as the previously used time harmonic potential. The main observation for the time harmonic modulation is the cloaking effect for the central Floquet side band in case of the Klein (i.e., for normal incidence of electron on the electrostatic barrier) transmission. On the other hand, for rectangular potential modulation in the time domain, the angular transmission profile exhibits asymmetric Fano resonances, a characteristic feature of the photon assisted tunneling. It is noted that the position and magnitude of the FR in the angular transmission spectra are sensitive both on the height and frequency of the rectangular modulation. Moreover, we demonstrate the features of the Fano resonances by adjusting ON and OFF phases of the modulating potential. Increasing the effective time of interaction over a period, the number of noted FR increases. We also note that there are some magic angles for which the transmission probabilities are protected against the change in the duty cycle. We conclude that the dressing of the barrier sharply depends on the height and duty cycle of the time dependent potential. It is shown that the angle averaged conductance through the tunneling structure can be effectively manipulated by tailoring the parameters of the rectangular potential. Thus, the use of rectangular time varying potential is superior over the harmonic one for the exploitation of the graphene n-p-n junction in the field of nanoscale engineering.

References:

[1] K. S. Novoselov, A. K. Geim, S. V. Morozov, D. Jiang, M. I. Katnelson, I. V. Grigorieva, S. V. Dubonos, and A. A. Firsov, Nature 438 (2005) 197.

[2] R. Mas-Balleste, C. Gomez-Navarro, J. Gomez-Herrero and F. Zamora, Nanoscale, 3 (2011) 20.

[3] Y. Lei et. al., ACS Nanosci. Au 2 (2022) 450.

[4] B. A. Salami, F. Mukhtar, S. A. Ganiyu, S. Adekunle and T. A Saleh, Construction and Building Materials 396 (2023) 132296.

[5] B. H. Nguyen and V. H. Nguyen, Adv. Nat. Sci: Nanotechnol. 7 (2016) 023002.

[6] E. Karaca and N. Acarali, Materials today communications 37 (2023) 107054.

[7] M. Asadi et. al., Journal of translational medicine 22 (2024) 611.

[8] N. Stander, B. Huard and D. Goldhaber-Gordon, Phys. Rev. Lett. 102 (2009) 026807.

[9] Z. Zhang et. al., Phys. Rev. Lett. 129 (2022) 233901.

[10] P.M. Ostrovsky et. al., Phys. Rev. B, 77(2008) 195430.

[11] M. I. Katsnelson, K. S. Novoselov, and A. K. Geim, Nature Phys. 2 (2006) 620.

[12] M. R. Masir, P. Vasilopoulos and F. M. Peeters, Phys. Rev. B, 82 (2010) 115417.

[13] O. V. Kibis, Phys. Rev. B 81 (2010) 165433.

[14] H. L. Calvo et. al., Appl. Phys. Lett 98 (2011) 232103.

[15] A. Iurov et. al., Phys. Rev. B 105 (2022) 115309.

[16] A. Iurov et. al., Phys. Rev. Reser. 2 (2020) 043245.

[17] C. Sinha and R. Biswas, Appl. Phys. Lett. 100 (2012) 183107.

[18] P. K. Tien and J. P. Gordon, Phys. Rev. 129 (1963) 647.

[19] W. Li and L.E. Reichl, Phys. Rev. B 60 (1999) 15732.

[20] M. Wagner et. al. 57 (1998) 11899.

[21] B. Trauzettel, Y. M. Blanter, A. F. Morpurgo, Phys. Rev. B 75 (2007) 035305.

[22] M. A. Zeb, K. Sabeeh and M. Tahir, Phys. Rev. B 78 (2008) 165420.

[23] R. Biswas, S. Maiti, S. Mukhopadhyay and C. Sinha, Phys. Letts. A 381 (2017) 1582.

[24] W. T. Lu et. al., J. Appl. Phys. 111 (2012) 103717.

[25] R. Biswas and C. Sinha, J. Appl. Phys. 115 (2014) 133705.

[26] R. Zhu, J.-H. Dai, and Y. Guo, J. Appl. Phys. 17 (2015) 164306.

[27] L. Z. Szabo, M. G. Benedict, A. Czirjak and P. Foldi, Phys. Rev B 88 (2013) 075438.

[28] A. Farrell and T. Pereg-Barnea, Phys. Rev. B 93 (2016) 045121.

[29] R. Biswas, S. Maiti, S. Mukhopadhyay and C. Sinha, Phys. Letts. A 381 (2017) 1582.

[30] R. Zhu and C. Cai, J. of Appl. Phys. 122 (2017) 124302.

[31] P. Mondal, S. Ghosh and M. Sharma, J. Phys. Condensed Matt. 31 (2019) 495001.

[32] O. Balabanov and H. Johannesson, J. Phys.: Condens. Matter 32 (2020) 015503.

[33] F. Sattari and S. Mirershadi, Phys. Scr. 95 (2020) 075702.

[34] Y. Betancur-Ocampo, P. Majari, D. Espitia F. Leyvraz and T. Stegmann, Phys. Rev. B 103 (2021) 155433.

[35] R. Biswas and C. Sinha, Scientific Reports 11 (2021) 2881.

[36] S. Bera, S. Sekh and I. Mandal, Ann. Phys. 535(2023) 2200460.

[37] J. Qin, H. Peng, J. Zhang and X. Hou, Appl. Phys. Lett. 109 (2016) 153303.

[38] J. Millman and C. C. Halkias, Integrated Electronics: Analog and Digital circuits and Systems, International student edition, McGraw-Hill Kogakusha LTD., 1972.

[39] J. Liu et. al., Vacuum 225 (2024) 113219.

[40] H. C. Shim, Y. K. Kwak, C. S. Han and S. Kim, Physica E 41 (2009) 1137.

[41] J. Saraswat and P. P. bhattacharya, Int. J. of comp. Net. And Commun. 5 (2013) 125.

[42] C. Sinha and R. Biswas, Appl. Phys. Lett. 100 (2012) 183107.

[43] R. Biswas and C. Sinha, Journal of Applied Physics 114 (2013) 183706.

[44] M. Büttiker, Phys. Rev. Lett. 57 (1986) 1761.

[45] R. Biswas, A. Biswas, N. Hui and C. Sinha, J. Appl. Phys. 108 (2010) 043708.

[46] S. Chen et. al. Science 353 (2016) 1522.

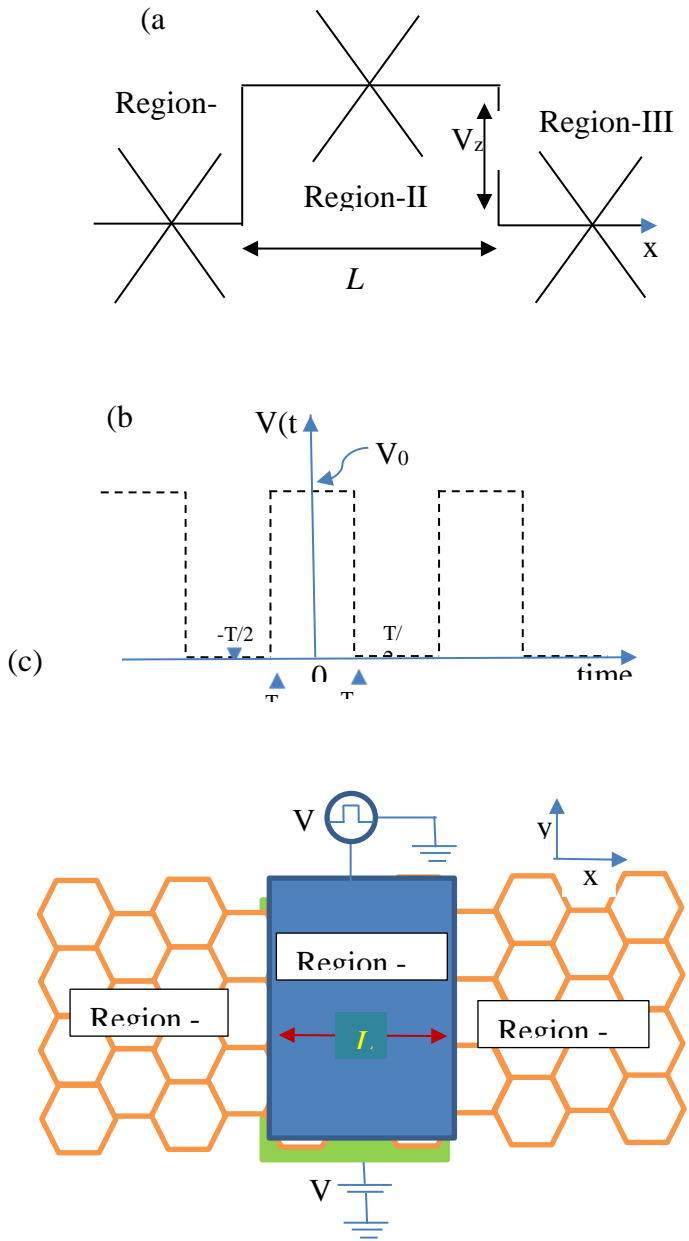

Fig.1: (a) Potential profile for monolayer graphene n-p-n structure. Potential is non uniform along the x-direction. L and $V_z$ correspond to the width and height of the electrostatic barrier (Region-II). (b) Potential profile for the time dependent rectangular potential of height $V_0$ and time period T applied in Region-II. (c) The schematic diagram of the proposed tunneling structure. Graphene sheet is in the x-y plane. Region I and Region II act as source and drain respectively. In Region-II electrostatic barrier is created by applying a back gate (green electrode) connected to a DC source $V_g$, whereas time dependent potential is applied by a top gate (blue electrode) connected to a multivibrator $V_t$.

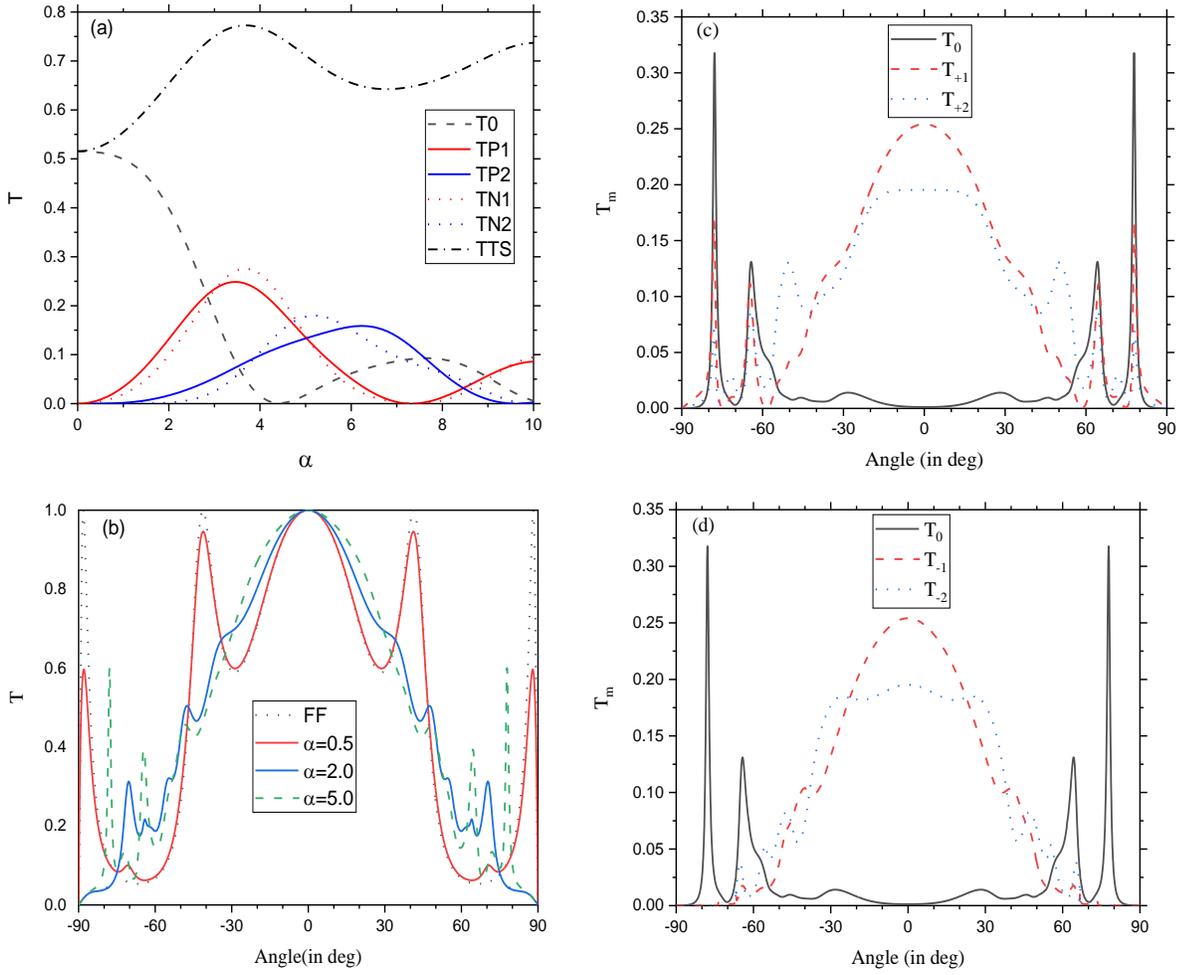

Fig.2: (Color online only) (a) Transmission coefficients $T_m$ for $m^{th}$ sideband in case of sinusoidal potential are plotted as a function of the parameter $\alpha \left(= \frac{V_0}{\omega}\right)$ of the periodic potential for energy E=10.21, static barrier height $V_z$=24.65, barrier width L=1.233 and angular frequency $\omega$=0.406. Angle of incidence $\theta = 30^0$. Dash-dot (black) line for total transmission (summed over all sidebands). (b) Total transmission for sinusoidal potential plotted as a function of the incidence angle. Dot (black) line for $\alpha = 0$ (Static barrier); solid (red) line for $\alpha = 0.5$; solid (blue) line for $\alpha = 2$ and dash (green) line for $\alpha = 5$. Other parameters are same as (a).

(c) Side band transmission coefficients for absorption channel are plotted as a function of incidence angle for $\alpha = 5.0$. (d) same as (c) but for emission channels.

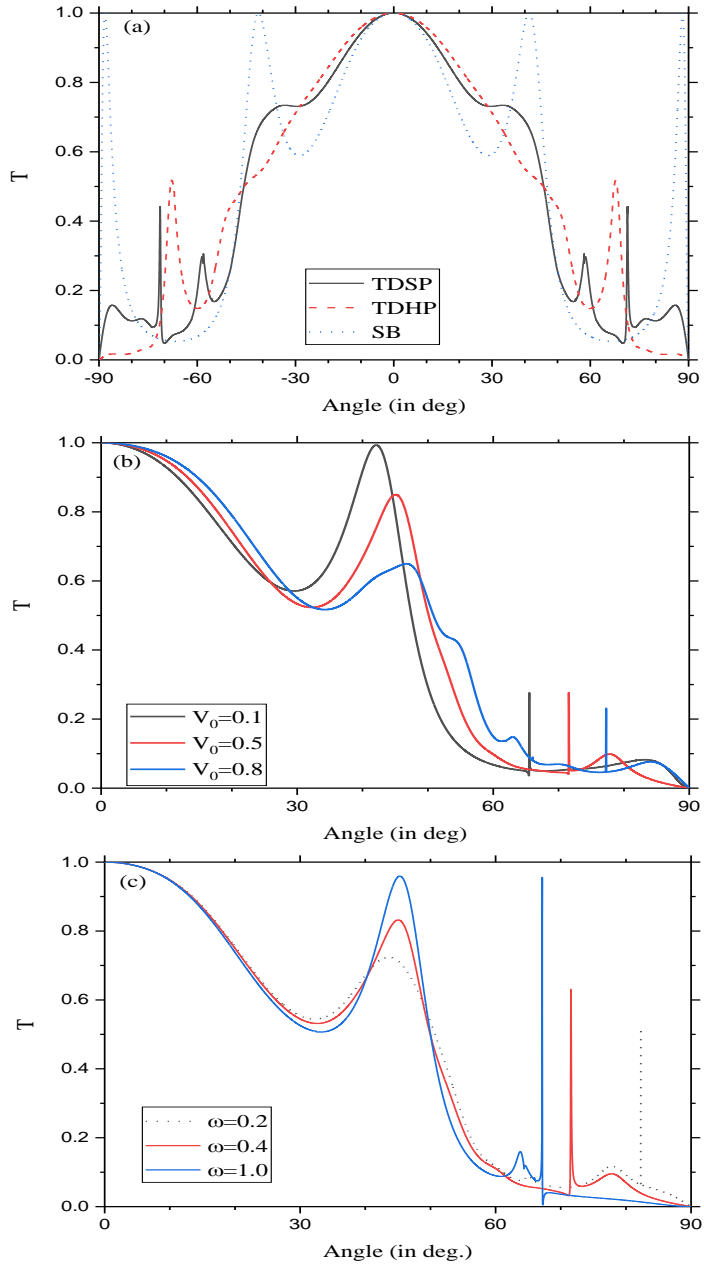

Fig.3: (Color online only) (a) Total transmission T (summed over all Floquet sidebands) is plotted as a function of angle of incidence for $V_0 = 0.5$ and others parameters are same as Fig.2(a). Solid (black) line for TDSP; Dash (red) line for TDHP and dot (blue) line for static barrier. (b) Same as (a) but only for TDSP. Solid (black) line for $V_0 = 0.1$; solid (red) line for $V_0 = 0.5$; solid (blue) line for $V_0 = 0.8$. (c) Same as (b) but for different values of frequency. Dot (black) line for $\omega = 0.2$; solid (red) line for $\omega = 0.4$; solid (blue) line for $\omega = 1$.

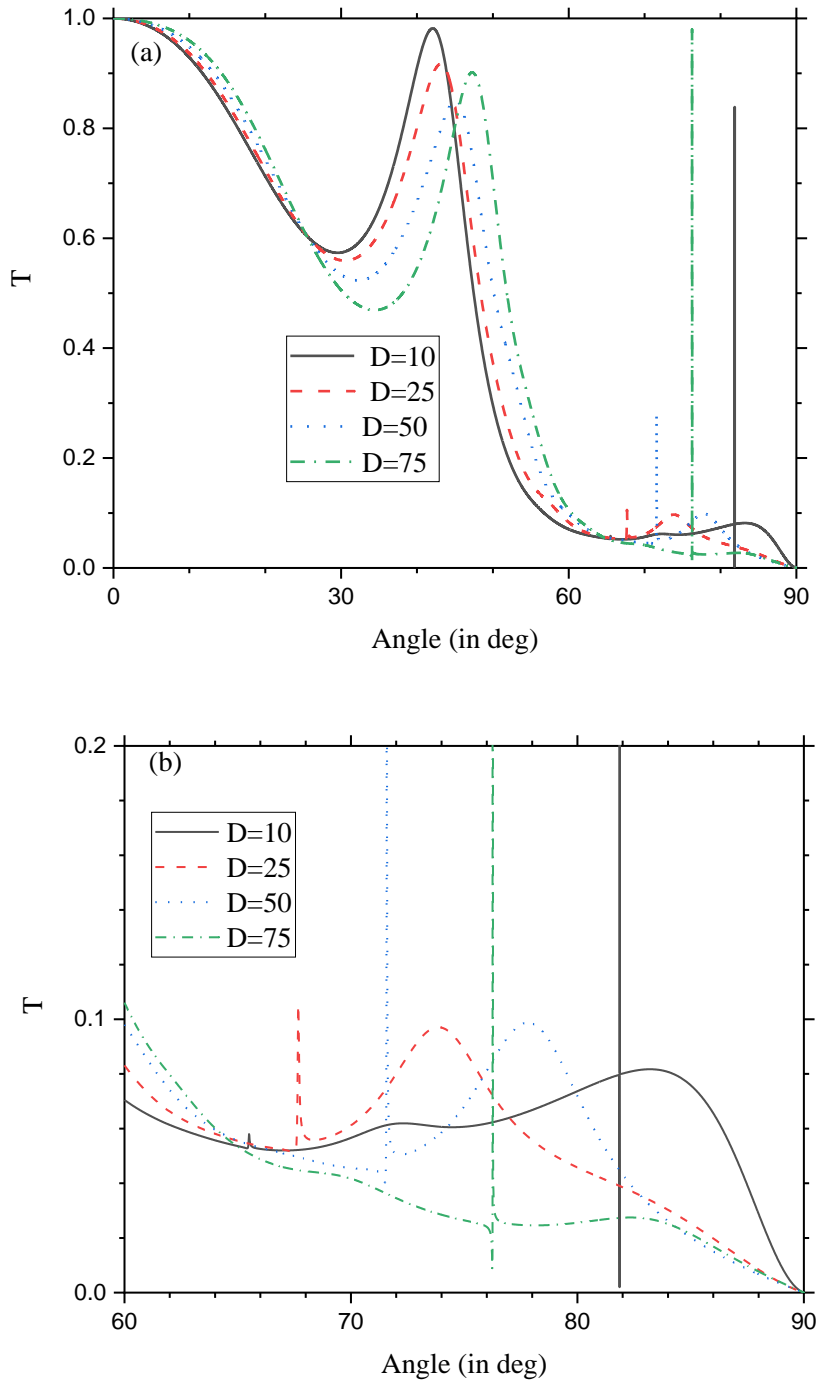

Fig.4: (Color online only) Same as Fig.2 but (a) for different values of the duty cycle (D). Solid (black) line for D=10%; dash (red) line for D=25%; dot (blue) line for D=50%; dot-dash (green) line for D=75%. (b) Extended part of the Fano resonance in fig.(a) for greater visibility.

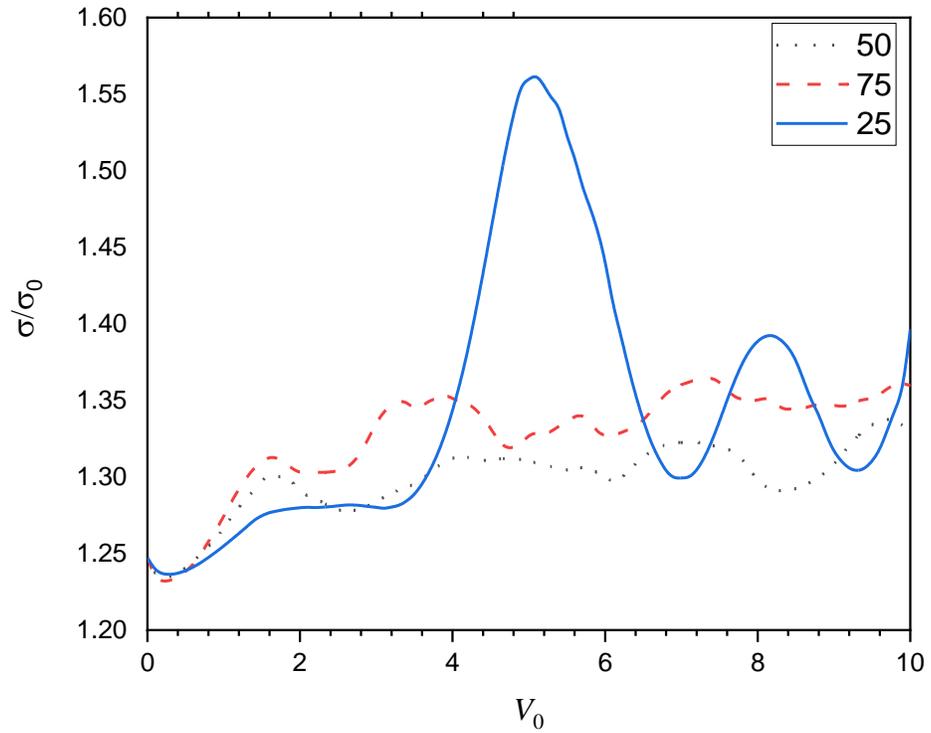

Fig.5: (Color online only) Angle averaged conductance σ (in units of σ$_0$) are plotted as a function of the height of the rectangular potential $V_0$. Solid (blue) line for D=25%; dotted (black) line for D=50% and dash (red) line for D=75%. Other parameters are same as Fig. 2(a).